\documentclass[letter]{aa}

\usepackage{txfonts}
\usepackage[dvips]{graphicx}
\usepackage{natbib}
\usepackage{multirow}

\def\EQ{\begin{equation}}
\def\EN{\end{equation}}
\def\EQA{\begin{eqnarray}}
\def\ENA{\end{eqnarray}}
\def\uu{{\bf u}}
\def\vv{{\bf v}}

\def\UU{{\bf U}}

\def\A{\mathcal{A}}

\def\OOmega{{\mbox{\boldmath$\Omega$}}}
\def\OB{\bar{\Omega}}

\begin{document}

\title{Effect of rotation on the tachoclinic transport}

\author{Nicolas Leprovost \and Eun-jin Kim}

\institute{Department of Applied Mathematics, University of Sheffield, Sheffield S3 7RH, UK}

\date{Received / Accepted }

\abstract{}{We study the effect of rotation on sheared turbulence due to differential rotation in the solar tachocline.}{By solving quasi-linear equations for the fluctuating fields, we derive turbulence amplitude and turbulent transport coefficients (turbulent viscosity and diffusivity), taking into account the effects of shear and rotation on turbulence. We focus on the regions of the tachocline near the equator and the poles where rotation and shear are perpendicular and parallel, respectively.}{For parameter values typical of the tachocline, we show that the shear reduces both turbulence amplitude and transport,  more strongly in the radial direction (parallel to the shear) than in the horizontal one, resulting in anisotropic turbulence.  Rotation further reduces turbulence amplitude and transport at the equator whereas it does not have much effect near the pole. The interaction between shear and rotation is shown to give rise to a novel non-diffusive flux of angular momentum (known as the $\Lambda$-effect), possibly offering a mechanism for the occurrence of a strong shear region in the solar interior. Further implications for the transport in the tachocline are discussed.}{}


\maketitle

\section{Introduction}
Solar rotation, both global and differential, plays a crucial role in the dynamical processes taking place in the Sun. For example, mean-field dynamo theory \citep{Moffatt78,Krause80} uses these two ingredients to explain the presence of the solar magnetic field. Firstly, the differential rotation in the tachocline is responsible for the in-situ generation of toroidal field via the $\Omega$ effect. Secondly, convection under the influence of (global) rotation leads to helical motion in the convection zone which permits the creation of poloidal fields via the $\alpha$ effect. 

To understand the differential rotation in the convective zone, many authors investigated the structure of turbulence in rotating bodies. The main feature of this type of turbulence is the appearance of non-diffusive term in the transport of angular momentum which prevents a solid body rotation from being a solution of the Reynolds equation \citep{Lebedinsky41,Kippenhahn63}. Starting from Navier-Stokes equation, it is possible to show that these fluxes arise when there is a cause of anisotropy in the system, either due to an anisotropic background turbulence \citep[see][and references therein]{Rudiger89} or due to inhomogeneity such as an underlying stratification \citep{Kichatinov87}. To explain the observed internal differential rotation and the depletion of light elements on the surface of the Sun, it is also of prime importance to understand the influence of rotation on turbulent transport coefficients such as the turbulent heat conductivity and the turbulent diffusivity of particles. 

The purpose of this Letter is to investigate the effect of (global) rotation on the tachocline transport. In our previous works, we have studied the turbulent transport in the tachocline by taking into account the crucial effect of shearing, the so-called shear stabilisation, due to a strong radial differential rotation \citep{Kim05,2Shears}. We have also incorporated the interaction of this sheared turbulence with different types of waves that can be excited in the Sun due to magnetic fields \citep{BetaPlane} or stratification \citep{Stratification}. In this Letter, we elucidate the effect of global rotation on sheared turbulence by studying a (local) Cartesian model valid near the equator and the poles. We consider a turbulence driven by an external forcing such as plumes from the convection zone and perform a quasi-linear analysis to derive the dependence of turbulence amplitude and transport on rotation and shear. Compared to two-dimensional turbulence studied in \citet{BetaPlane}, global rotation supports the propagation of inertial waves in three dimensions (3D), which interact with the shear flow, playing an important role in the overall turbulent transport. In particular, we report a novel result that the momentum transport can  be not only due to eddy viscosity but also to non-diffusive $\Lambda$-effect in the tachocline. Non trivial $\Lambda$-effect here results from an anisotropy induced by shear flow on the turbulence even when the driving force is isotropic, in contrast to the case without shear flow where this effect exists only for anisotropic turbulence \citep{Kichatinov87}.

\section{Model}
\label{Model}
\begin{figure}[h]
\begin{center}
\includegraphics[scale=1.2,clip]{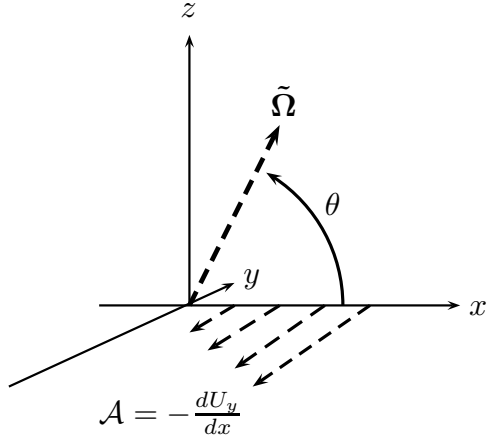}
\end{center}
\caption{\label{DessinEq} The configuration in our model: $\A$, $\tilde{\Omega}$ and $\theta$ are the shearing rate, the rotation rate and the co-latitude, respectively.}
\end{figure}
To elucidate the effect of rotation on sheared turbulence, we model the tachocline by an incompressible fluid in local Cartesian coordinates $(x,y,z)$ in a rotating frame (see Fig. \ref{DessinEq}) with average rotation rate $\tilde{\Omega}$ making an angle $\theta$ with a mean shear flow in the azimuthal direction: $\UU_0  = - x \mathcal{A} \hat{j}$. We study the effect of this large-scale shear on the transport properties of turbulence by assuming the velocity as a sum of a radial shear (i.e. in the $x$-direction) and fluctuations: $\uu = \UU_0 + \vv = U_0(x) \hat{j} + \vv = - x \mathcal{A} \hat{j} + \vv$. We resort to the quasi-linear approximation \citep{Townsend76,Moffatt78}, also called first-order smoothing (FOSA) or second-order correlation approximation (SOCA), where the product of fluctuations is neglected to obtain the following equations for the evolution of the fluctuating velocity field: 
\EQA
\label{quasi-linear}
\partial_t \vv + \UU_0 \cdot \nabla \vv + \vv \cdot \nabla \UU_0  &=& - \nabla p + \nu \nabla^2 \vv + {\bf f} - \OOmega \times \vv \; , \\ \nonumber
\nabla \cdot \vv &=& 0 \; , 
\ENA
where $p$ and ${\bf f}$ are respectively the small-scale components of the pressure and forcing, and ${\bf \Omega} \equiv 2 \tilde{\bf \Omega}$. 

To study the influence of rotation and shear on the particle and heat transport, we have to supplement Eq.~(\ref{quasi-linear}) with an advection-diffusion equation for these quantities. We here focus on the transport of particles since a similar result also holds for the heat transport. The density of particles is also written as a sum of a large-scale component $N_0$ and fluctuations $n$. Using the quasi-linear approximation, the fluctuating concentration of particles is governed by the following equation:
\EQ
\label{EquTransport}
\partial_t n + \UU_0 \cdot \nabla n + \vv \cdot \nabla N_0 = D \nabla^2 n \; ,
\EN
where $D$ is the molecular diffusivity of particle. Note that, in the case of heat equation, $D$ should be replaced by the molecular heat conductivity $\chi$. 

As the large-scale velocity is in the $y$ (azimuthal) direction, we are mostly interested in the momentum transport in that direction. The equation for the (large-scale) azimuthal velocity $U_0$ is then given by Navier-Stokes equation with the contribution from fluctuations given by ${\bf \nabla \cdot R}$, where ${\bf R} = \langle {\bf v} \mathrm{v}_y \rangle$ is the Reynolds stress. One can formally Taylor expand ${\bf R}$ with respect to the gradient of the large-scale flow:
\EQ
\label{BullShit200}
R_i = \Lambda_i U_0 - \nu_T \partial_x U_0 \, \delta_{i1} + \dots = \Lambda_i U_0 + \nu_T \A \, \delta_{i1} + \dots \; .
\EN
where we introduced two coefficients $\Lambda_i$ and $\nu_T$. The effect of the turbulent viscosity $\nu_T$ is simply to change the viscosity from the molecular value $\nu$ to the effective value $\nu + \nu_T$. In comparison, the first term $\Lambda_i U_0$ in Eq.~(\ref{BullShit200}) is proportional to the velocity rather than its gradient. This means that it does not vanish for a constant velocity field and can thus lead to the creation of gradient in the velocity field. This term is equivalent to the $\alpha$-effect in dynamo theory \citep{Parker55,Steenbeck66b}, and has been known as the $\Lambda$-effect \citep{Krause74} or anisotropic kinetic alpha (AKA)-effect \citep{Frisch87}. Similarly, the transport of species in the large scales is governed by an advection diffusion equation where the molecular diffusivity is supplemented by a turbulent diffusivity, defined as $\langle \mathrm{v}_i n \rangle = - D_T^{ij} \partial_j N_0$. 

To calculate the turbulence amplitude and transport coefficients (Reynolds stress and turbulent diffusivity), we prescribe the forcing in Eq.~(\ref{quasi-linear}) to be incompressible, isotropic and short correlated in time (modelled by a  $\delta$-function) with a power spectrum $F$. Specifically, we assume:
\EQA
\label{ForcingCorrel}
\langle \tilde{f}_i({\bf k}_1,t_1) \tilde{f}_j({\bf k}_2,t_2) \rangle &=& \tau_f \, (2\pi)^3 \delta({\bf k}_1+{\bf k}_2) \, \delta(t_1-t_2) \times \\ \nonumber
&& \qquad F(k) \, (\delta_{ij} - k_i k_j/k^2) \; .
\ENA
The angular brackets stand for an average over realisations of the forcing, and $\tau_f$ is the (short) correlation time of the forcing. 
We solve Eqs.~(\ref{quasi-linear}) and  (\ref{EquTransport}) in the case of unit Prandtl number ($\nu = D$) for simplicity, and use the results of these and Eq.~(\ref{ForcingCorrel}) to compute the turbulent transport. 
In this Letter, we focus on the case relevant to the Sun and discuss the implications of our findings for the dynamics of the tachocline. Note that in the tachocline, $\vert \A \vert \sim 3 \times 10^{-6} \, \mbox{s}^{-1}$ by using the upper limit on the tachocline thickness ($5\%$ of the solar radius) while the average rotation rate of the interior is $\tilde{\Omega} \sim 2.6 \times 10^{-6} \, \mbox{s}^{-1}$. Therefore, the ratio of rotation to shear is very close to unity. However, if the tachocline is found to be thinner, the value of the shearing rate $\A$ becomes larger, with $\vert \Omega  / \A \vert < 1$. Furthermore, with molecular viscosity of $\nu \sim 10^2 \, \mbox{cm}^2\mbox{s}^{-1}$,  $\xi = \nu k_y^2 / \vert \A \vert$ is a small parameter for a broad range of reasonable length scales in the azimuthal direction, $L_y > 10^4 \mbox{cm}$. Thus, the present results are valid for slow rotation ($\Omega \ll \vert \A \vert$) and strong shear ($\xi \ll 1$). 

\section{Near the equator ($\theta = \pi / 2$)}
Near the equator, the rotation and the direction of the shear are orthogonal. Expanding the velocity in powers of $\OB = \Omega / \A $, we compute the turbulence amplitude and transport coefficients up to second order in  $\vert \OB \vert \ll 1$.
\subsection{Turbulence intensity}
\label{Intensity}
In the strong shear limit ($\xi \ll 1$), using similar algebra as in \citet{Kim05}, we obtain the turbulent intensity in the radial direction as follows:
\EQ
\label{Vxweak}
\langle \mathrm{v}_x^2 \rangle = \frac{\tau_f}{ \A} \int \frac{d^3 k}{(2 \pi)^3} \,  F(k) \left[L_0({\bf k}) + \OB  \frac{k_z^2}{k_y^2} L_1({\bf k})\right] \; .
\EN
Here, $k_\mathrm{H}^2 = k_y^2 + k_z^2$ is the (squared) amplitude of the wave number in the horizontal plane, and $L_0$ and $L_1$ are two positive definite integrals depending only on the wave number ${\bf k}$. Note that the details of $L_0$ and $L_1$ are not important for our discussions. Therefore, the turbulence intensity $\langle \mathrm{v}_x^2 \rangle$ in Eq.~(\ref{Vxweak}) increases for $\OB > 0$ whereas it decreases for $\OB < 0$. This is because a weak rotation destabilises sheared turbulence for $\OB > 0$ whereas it stabilises for $\OB < 0$ \citep{Bradshaw69,Salhi97}.

For the other components of the turbulence amplitude, we obtain:
\EQA
\label{Vzweak}
\langle \mathrm{v}_z^2  \rangle &\sim& \frac{\tau_f}{ \A} \int \frac{d^3 k}{(2 \pi)^3}  \, F(k) L_2({\bf k}) \left(\frac{3}{2\xi}\right)^{1/3} \times  \\ \nonumber 
&& \qquad \left[\Gamma(1/3) + \OB \frac{k_z^2}{k_y^2} \Gamma(4/3) (-\ln\xi)\right] \; .
\ENA
Here, $\Gamma$ is the Gamma function and $L_2$ is a positive definite function of ${\bf k}$. Here again, the turbulence amplitude can increase or decrease depending on the sign of $\OB$. Furthermore, the correction due to rotation has a logarithmic dependence on the shear, contrary to the case of the radial amplitude. Note that the anisotropy in the first correction is less pronounced than in the leading order term. As a result, the turbulence due to shearing becomes less anisotropic. This illustrates the tendency of rotation to lead to almost isotropic turbulence.  An equation similar to (\ref{Vzweak}) is also found for $\langle \mathrm{v}_y^2 \rangle$ with a slightly different function $L_2$. However, performing the angular integration, we find that $\langle \mathrm{v}_y^2 \rangle$ is larger than $\langle \mathrm{v}_z^2 \rangle$, in agreement with numerical simulations \citep{Lee90}. This effect is present for $\OB = 0$ and is thus caused by the shear only. 

\subsection{Transport of angular momentum}
In the strong shear limit ($\xi \ll  1$), the transport of angular momentum can be found as a sum of two terms $\langle \mathrm{v}_x \mathrm{v}_y \rangle = \nu_T \A + \Lambda_x \Omega$, the first and second  being  even and odd with the rotation rate, respectively. The first term, the turbulent viscosity, takes the following form:
\EQ
\label{TurbViscos}
\nu_T \sim \frac{\tau_f}{ \A^2} \int \frac{d^3 k}{(2 \pi)^3}  \, F(k) \left[-\frac{1}{2} +  L_3({\bf k}) \right] \; ,
\EN
where $L_3$ is a positive definite function of ${\bf k}$. For $\OB = 0$ we recover the result of \citet{Kim05} showing that the turbulent viscosity is reduced proportionally to $\A^{-2}$ for strong shear. The turbulent viscosity  can be either positive or negative depending on the relative magnitude of the two terms inside the integral. In the 2D limit (where $L_3=0$), we can easily see that the turbulent viscosity is negative. On the contrary, in the case of an isotropic forcing in 3D, the turbulent viscosity is positive.

The correction due to the rotation is proportional to $\Omega$ and is thus odd in the rotation. This is the so-called $\Lambda$-effect, a non-diffusive contribution to Reynolds stress, which can be shown to be:
\EQ
\label{LambdaEquator}
\Lambda_x \sim \frac{\tau_f}{ \A^2} \int \frac{d^3 k}{(2 \pi)^3}  \, F(k) L_4({\bf k}) \, (-\ln\xi) \; .
\EN
It is important to emphasise that this non trivial $\Lambda$-effect results from an anisotropy induced by shear flow on the turbulence even when the driving force is isotropic. This should be contrasted to the case without shear flow where non-diffusive fluxes emerge only for anisotropic forcing \citep{Rudiger80,Kichatinov86b}. In our case, the anisotropy in the velocity field is not artificially introduced in the system but is created by the shear and calculated self-consistently.

\subsection{Transport of particles}
In the strong shear limit ($\xi \ll 1$), the computation of $\langle n {\bf v} \rangle$ gives:
\EQA
\label{TransportEquatWeak}
D_T^{xx} &\sim& \frac{\tau_f}{ \A^2} \int \frac{d^3 k}{(2 \pi)^3}  \, F(k) L_5({\bf k}) \left[1+\OB \frac{k_z^2}{k_y^2} \,  \frac{-\ln\xi}{3} \right] \; , \\  
\label{TransportEquatWeak2}
D_T^{zz} &\sim& \frac{\tau_f}{\A^2} \int \frac{d^3 k}{(2 \pi)^3} F(k) L_6({\bf k}) \left(\frac{3}{2\xi}\right)^{\frac{2}{3}}  \left[1 + 2 \OB \frac{k_z^2}{k_y^2} \,  \frac{-\ln\xi}{3} \right]  \; ,
\ENA
where $L_5$ and $L_6$ are positive definite functions.  Eqs.~(\ref{TransportEquatWeak}-\ref{TransportEquatWeak2}) show that the effect of rotation on the transport of particles depends on the sign of $\OB$: for $\OB > 0$, the transport is increased whereas it is reduced when $\OB < 0$. This is again because a weak rotation destabilises sheared turbulence for $\OB > 0$ whereas it stabilises for $\OB < 0$. Note that a similar behaviour was also found in turbulence intensity, given in Eqs.~(\ref{Vxweak}) and (\ref{Vzweak}). The correction term due to rotation in $D_T^{xx}$ and $D_T^{zz}$ in Eq.~(\ref{TransportEquatWeak}-\ref{TransportEquatWeak2}) depends weakly on the shear by a logarithmic factor $\vert \ln \xi \vert$, which cannot be too large even for $\xi \ll 1$, and is of the same order for transport in different directions. Thus, the scaling of the turbulent diffusivity is roughly the same as that in the case without rotation: the radial transport ($\propto \A^{-2}$) is more reduced than the horizontal one ($ \propto \A^{-4/3}$). This result should be contrasted to the rapid rotation limit where the transport in the radial direction was larger (but only by a factor $2$) than the one in the horizontal direction \citep{Rudiger89}. These results thus highlight the crucial role of shear in transport, in particular, by introducing anisotropy.

\section{Near the poles ($\theta = 0$)}
Near the poles, the directions of rotation and shear can be taken to be parallel. Contrary to the case at the equator, we find that the leading order correction (proportional to $\OB$) vanishes in the case of an isotropic forcing (due to the fact that these terms are odd in one of the components of the wave number) for all the previously calculated quantities. Consequently, the turbulent amplitude, viscosity and diffusivity are  given by Eqs.~(\ref{Vxweak}-\ref{Vzweak}), (\ref{TurbViscos}) and (\ref{TransportEquatWeak}-\ref{TransportEquatWeak2}) with $\OB =0$, respectively. Similarly, the Reynolds stress involving the radial component ($\langle \mathrm{v}_x \mathrm{v}_y \rangle$) vanishes.  However the component of the Reynolds stress involving the latitudinal velocity ($\langle \mathrm{v}_y \mathrm{v}_z \rangle$) does not vanish and is odd in $\Omega$. Thus, the $\Lambda$-effect appears here in  $\langle \mathrm{v}_y \mathrm{v}_z \rangle$ (recall that at the equator, the $\Lambda$-effect was present only in $\langle \mathrm{v}_x \mathrm{v}_y \rangle$), and takes the following form:
\EQ
\label{LambdaPole}
\Lambda_z \sim - \frac{\tau_f}{ \A^2} \int \frac{d^3 k}{(2 \pi)^3} \,  F(k) \left(\frac{3}{2\xi}\right)^{2/3} \left[L_7({\bf k}) - L_8({\bf k}) \right]\; .
\EN
Eq.~(\ref{LambdaPole}) shows that $\Lambda_z$ is of indefinite sign,  as $L_7$ and $L_8$ are two positive definite functions but appear with different signs. However, as for Eq.~(\ref{LambdaEquator}), in the case of an isotropic forcing, the function $L_7$ dominates over $L_8$, leading to a negative $\Lambda_z$. This $\Lambda$-effect arises even in the case of an isotropic forcing as the shear favours fluctuations in the $y$-direction compared to that in the $z$-direction (see \S \ref{Intensity}), leading to anisotropic turbulence in the plane perpendicular to the rotation. Eq.~(\ref{LambdaPole}) also shows that $\Lambda_z$ is larger (scaling as $\A^{-4/3}$) than $\Lambda_x$ in the equatorial case.

\section{Implications for the tachocline}
\label{Implications}
Table \ref{Summary} summarises the scaling of our results with the shear $\A$ and rotation $\Omega$ in the limit of weak rotation ($\Omega \ll \vert \A \vert$) and strong shear ($\xi = \nu k_y^2 / \vert \A \vert \ll 1$).
\renewcommand{\arraystretch}{1.5}
\begin{table}[h]
\begin{tabular}{|c|c|c|}
\hline
& Equator & Pole \\ \hline
$\langle \mathrm{v}_x^2 \rangle$ & $\A^{-1} \left[1 + c \OB\right]$ & $\A^{-1} $ \\ \hline
$\langle \mathrm{v}_y^2 \rangle \sim \langle \mathrm{v}_z^2 \rangle$ & $\A^{-2/3} \left[1 + c \OB \vert \ln \xi \vert\right]$ & $\A^{-2/3}$ \\ \hline
$ \nu_T $ & $\A^{-2} $ & $\A^{-2} $ \\ \hline
$\Lambda_x $ & $\A^{-2} \vert \ln \xi \vert$ & $\qquad$ 0 $\qquad$  \\ \hline
$\Lambda_z $ &  0 & $- \A^{-4/3}$ \\ \hline
$D_T^{xx}$ & $\A^{-2} \left[1 + c \OB \vert \ln \xi \vert\right]$ & $\A^{-2} $ \\ \hline
$D_T^{yy} \sim D_T^{zz}$ & $\A^{-4/3} \left[1 + c \OB \vert \ln \xi \vert\right]$ & $\A^{-4/3} $  \\ \hline
\end{tabular}
\caption{\label{Summary} Scaling of turbulence amplitude, turbulent viscosity ($\nu_T$), $\Lambda$-effect and turbulent diffusivity with the shear $\A$ and rotation $\OB = \Omega / \A$. $c$ is a positive constant of order unity.}
\end{table}

The first two rows show that, both near the equator and poles, the turbulence amplitude in the radial direction is more reduced by the shear than in the horizontal one, by a factor of  $\A^{-1}$ and $\A^{-2/3}$, respectively. These results thus imply an effectively stronger turbulence in the horizontal ($y$-$z$) plane than the one in the radial ($x$) direction. Furthermore, at the equator, $\OB < 0$ as the shear increases from the interior towards the convective zone. Thus, the turbulence amplitude is further reduced by the rotation in all directions, but the turbulence in the radial direction is more suppressed than the one in the horizontal direction by a factor $\ln\xi$. 

A novel result is also found for the Reynolds stress, which involves two contributions: the turbulent viscosity and the $\Lambda$-effect. The latter, a source of non-diffusive flux, is present even with isotropic forcing due to the shear-induced anisotropy \citep{Kim05}. Note that for an isotropic forcing, the turbulent viscosity is found to be positive and therefore cannot act on its own as a source of differential rotation in the tachocline. The radial differential rotation can, however, be created by the $\Lambda$-effect: since it does not depend only on the gradient of angular velocity, it does not vanish for a uniform rotation and thus prevents the uniform rotation from being a solution of the large-scale momentum equation. To determine the sign of $\A$ as a result of the $\Lambda$-effect, we seek a solution of the large-scale turbulent equations by demanding the Reynolds stress to vanish \citep[e.g.][]{Kichatinov87}. In the equatorial case, we then obtain the following equation for the shear:
\EQ
\nu_T(\A) \A + \Lambda(\A) \Omega = 0 \; .
\EN
As both $\nu_T$ and $\Lambda$ are positive, $\A$ must be negative. Interestingly, this is exactly what the observation indicates (a rotation rate decreasing towards the interior at the equator). We cannot use such a simple equation to predict the sign of the shear at the pole as the $\Lambda$-effect now appears in the $\langle \mathrm{v}_y \mathrm{v}_z \rangle$ component of the Reynolds stress and is found to scale as $\A^{-4/3}$, which is larger than $\Lambda_x$ ($\propto \A^{-2}$) near the equator.

Furthermore, we find that the transport of chemical species is reduced by shear and  more severely in the the radial direction than the horizontal one (by a factor $\A^{-2}$ and $\A^{-4/3}$ respectively), with a further, but weak reduction due to rotation at the equator only. As a result, the radial mixing (in the $x$ direction) of chemicals is slightly more efficient at the pole than at the equator. These results thus suggest that turbulent mixing of light elements is weak (due to shearing) in the tachocline and also that the depletion of light elements on the solar tachocline depends on latitude. However, this latitudinal dependency is unlikely to be observed on the surface due to the rapid mixing in the convection zone \citep{Spruit77}. Since a similar result applies to heat transport, a faster heat transport at the pole could contribute to making hotter poles than equator at the solar surface. Finally, we note that all our results here are valid in the absence of stratification and magnetic fields \citep{Stratification} and will extend our study to include these in future publications. Of particular interest would be the sign of $\nu_T$ and $\Lambda$, and thus the resulting sign of $\A$. 

\begin{acknowledgements}
This work was supported by U.K. PPARC Grant No. PP/B501512/1.
\end{acknowledgements}

\bibliographystyle{aa}
\bibliography{Bib_RotShearAA}

\end{document}